\documentclass[12pt]{article}

\usepackage{amsfonts,amsmath,amssymb}

\newcommand{\Author}[2]{#1. #2}

\begin{document}

\title{
  There is no mass squared term
  \\
  in neutrino electric charge
}

\author{
  Maxim Dvornikov\thanks{E-mail: maxim\_dvornikov@aport.ru}
  \and
  Alexander Studenikin\thanks{E-mail: studenik@srd.sinp.msu.ru}
  \\
  Department of Theoretical Physics, Moscow State University,
  \\
  119992 Moscow, Russia
}

\date{}

\maketitle

\begin{abstract}
We claim that the neutrino mass squared term cannot appear in the
expansion of the neutrino electric charge over the neutrino mass
parameter at the one-loop level of the standard model supplied
with $\mathrm{SU}(2)$-singlet right-handed neutrino.
\end{abstract}

Neutrino electromagnetic properties is an interesting and
longstanding problem. The most important static electromagnetic
characteristics of neutrino are the electric charge and magnetic
moment. It is obvious that a massless particle can possess neither
electric charge nor magnetic moment. However, in the case of
non-zero mass, due to the radiative corrections the neutrino
static electromagnetic moments can have non-vanishing values.

Recently the massive neutrino electric charge $Q_\nu$ was
estimated in \cite{Sha03} as follows,
\begin{equation}
  \label{Qwrong}
  Q_\nu=-\frac{3eG_Fm_\nu^2}{4\pi^2\sqrt{2}},
  \quad
  e=-|e|,
\end{equation}
so it was supposed that the non-vanishing neutrino electric charge
could be proportional to the neutrino mass squared.

In this short note we argue that the estimation for the neutrino
electric charge, given by eq.(\ref{Qwrong}), is not correct. This
statement is based on our investigations \cite{DvoStu03} of the
massive neutrino electromagnetic form factors. In particular, we
calculated the one-loop contributions to the massive neutrino
electric charge and magnetic moment within the context of the
standard model supplied with $\mathrm{SU}(2)$-singlet right-handed
neutrino in arbitrary $R_{\xi}$ gauge. It is important for the
present discussion that our recent studies \cite{DvoStu03} enables
us to consider the dependence of the neutrino charge on the mass
of neutrino. The neutrino electric charge can be represented as a
series-expansion over the neutrino mass parameter
$b=(m_\nu/M_W)^2$,
\begin{equation}
  \label{Qexp}
  Q_\nu(a,b,\alpha)=Q_0(a,\alpha)+bQ_1(a,\alpha)+\mathcal{O}(b^2),
\end{equation}
where $m_\nu$, $M_W$, and $m_e$ are the neutrino, $W$ boson, and
charged lepton masses, respectively, and  $\alpha$ is the $W$
boson gauge parameter. In \cite{DvoStu03} it was shown by the
direct calculations that the both functions in eq.~(\ref{Qexp}),
$Q_0(a,\alpha)$ and $Q_1(a,\alpha)$,
  vanish for arbitrary values of
$a$ and $\alpha$. Thus, at the one-loop level there are no mass
squared contributions to the neutrino electric charge for any
gauge.

Moreover, we also shown in our paper \cite{DvoStu03} that for a
particular choice of the gauge fixing parameter $\alpha=1$ (the
't~Hooft-Feynman gauge), the neutrino electric charge is exactly
zero,
\begin{equation}
Q_\nu(a,b,\alpha=1)=0,
 \end{equation}
for any values of the mass parameters $a$ and $b$. It is obvious
again that there is no room for the mass squared terms in the
neutrino electric charge for arbitrary neutrino mass on the
one-loop level.

In the paper \cite{Sha03} there is an attempt to establish the
relation between the neutrino neutrality and masslessness on the
basis of eq.~(\ref{Qwrong}) using the results of
\cite{CabBerVidZep00}. Indeed, it was shown in
\cite{CabBerVidZep00} that the neutrino electric charge is
vanishing. However, in the study of the neutrino electric charge
in \cite{CabBerVidZep00} it was assumed just in the beginning of
the calculations that  the mass of neutrino is zero. Thus,
contrary to the last claim of \cite{Sha03}, the present limits 
on the neutrino mass and possible 
neutrino non-zero electric charge does not provide any
reason to introduce a new structure for electromagnetic gauge
invariance .

\end{document}